\documentclass[sigconf]{acmart}

\usepackage{amsfonts}
\usepackage{chemarrow}
\usepackage{hyperref}
\usepackage{amsmath} 
\usepackage{graphicx}
\usepackage{subfigure}
\usepackage{url}
\usepackage{color}
\usepackage{breqn}
\newcommand{\BfPara}[1]{\vspace{1mm}{\noindent\bf#1.}\xspace}
\usepackage{multirow}
\usepackage{float}
\usepackage{booktabs}
\usepackage{balance}

\usepackage{xspace,lipsum}

\newcommand{\etc}{{etc.}\xspace}
\newcommand{\eg}{{\em e.g.}\xspace}
\newcommand{\ie}{{\em i.e.,}\xspace}

\usepackage{makecell}

\usepackage{xcolor,colortbl}

\definecolor{darkgreen}{rgb}{0.0, 0.2, 0.13}
\definecolor{darkred}{rgb}{0.2, 0.0, 0.13}

\newcommand{\etal}{{\em et al.}\xspace}

\hypersetup{
	pdfpagemode=pagewidth,
	plainpages=false,
	colorlinks,
	urlcolor=blue,
	linkcolor=blue,
	citecolor=blue,
	bookmarksnumbered   
}

\newcommand{\ours}{{ShellCore}\xspace}

\colorlet{numb}{magenta!60!black}

\setcopyright{none}
\renewcommand\footnotetextcopyrightpermission[1]{}
\begin{document}

\title{{\sc ShellCore}: Automating Malicious IoT Software Detection by Using Shell Commands Representation}

\author{Hisham Alasmary$^{1,2}$, Afsah Anwar$^1$, Ahmed Abusnaina$^1$, Abdulrahman Alabduljabbar$^1$, Mohammad Abuhamad$^3$, An Wang$^4$, DaeHun Nyang$^5$, Amro Awad$^6$, David Mohaisen$^1$
}
\affiliation{$^1$University of Central Florida \hspace{2mm}$^2$King Khalid University $^3$Loyola University \hspace{2mm}$^4$Case Western Reserve  University \hspace{2mm}$^5$Ewha Womans University \hspace{2mm}$^6$NCSU}

\begin{abstract}
The Linux shell is a command-line interpreter that provides users with a command interface to the operating system, allowing them to perform a variety of functions. Although very useful in building capabilities at the edge, the Linux shell can be exploited, giving adversaries a prime opportunity to use them for malicious activities. With access to IoT devices, malware authors can abuse the Linux shell of those devices to propagate infections and launch large-scale attacks, e.g., DDoS.  In this work, we provide a first look at shell commands used in Linux-based IoT malware towards detection. We analyze malicious shell commands found in IoT malware and build a neural network-based model, \ours, to detect malicious shell commands. Namely, we collected a large dataset of shell commands, including malicious commands extracted from 2,891 IoT malware samples and benign commands collected from real-world network traffic analysis and volunteered data from Linux users. 
Using conventional machine and deep learning-based approaches trained with term- and character-level features, \ours is shown to achieve an accuracy of more than 99\% in detecting malicious shell commands and files (\ie binaries). 
\end{abstract}

\ccsdesc{Security and Privacy~Distributed systems security}



\maketitle

\section{Introduction}\label{sec:introduction}

Internet of Things (IoT) manufacturers and application developers have started to discover the benefits of the edge computing paradigm and do more compute and analytics on the devices themselves.
 The on-device approaches help reduce latency for critical applications, lower dependence on the cloud, and better manage the massive data generated by the IoT devices.
 An example of this trend is the Nest Cam IQ indoor security camera~\cite{NestCamIQIndoor}, which uses on-device vision processing power to watch for motion, distinguish family members, and send alerts. 
 Such a paradigm provides new opportunities for IoT applications~\cite{wolskiKBGL19,jangLSL18}. To unleash the power of Linux-based systems, IoT devices at the edge employ shell commands, which would allow invocation of Linux capabilities in a seamless manner. This utilization, which is essential for many edge applications, is sometimes exploited by malicious actors (malactors) to launch malicious activities, and automate the process of attacks and malware proliferation. 

Indeed, the increasing use of IoT devices for everyday activities has been paralleled with IoT's susceptibility to risks, including major attack vectors, such as vulnerabilities in the hardware and software stacks and the use of default usernames and passwords.
Those attack vectors are demonstrated by major high bandwidth Distributed Denial of Service (DDoS) attacks. Targets of those attacks include large companies, such as {\em Github}~\cite{GithHub1Tb} and {\em Dyn}~\cite{DynAttack}. 
To launch those attacks, the attackers exploit infected IoT devices for executing a series of commands for malware and attack propagation.
Since most IoT and embedded devices use a packed version of software, such as Busybox~\cite{wells2000}, to implement Linux capabilities, Linux-based shell commands are used for automating those attacks.

The Linux shell as an entry point to IoT devices is accessible to many attacks, including brute-force, privilege escalation, shellshock, and other vulnerabilities (\eg, CVE-2018-9310, CVE-2019-1656, CVE-2018-0183, CVE-2017-6707) ~\cite{nvd_cvss,MadeOfBug,ChenMWZZK11,UittoRML15}. Using secondary information, such as the listings of IoT and embedded devices on the likes of Shodan~\cite{Matherly09}, adversaries can utilize default passwords to connect to arbitrary devices on the Internet, gain control over them, and use them for their malicious activities through remote access and automation tools. For example, a simple ``default password'' search on Shodan returns 72,763 results, which all can be accessed, and used for attacks. 

Shell commands are heavily utilized in IoT malware and botnet operation. 
Malware-infected hosts use Command and Control (C2) servers to obtain payloads that include instructions to compromised machines (or bots). Such instructions aim to synchronize actions and cycles of activities to attack targets and propagate the recruitment of new bots that eventually become a source of propagation.
In this example, bots use the shell to execute {\em chmod} command to change privileges. Moreover, bots also use the shell to launch a dictionary brute-force attack and to propagate by connecting to the C2 server to download instructions using the HTTP protocols. 
To launch an attack, a bot typically obtains a set of targets from a dropzone by invoking a set of commands
that uses the shell to flood the HTTP of the victim and to remove the traces of execution by executing the {\em rm} command~\cite{AntonakakisABBB17}. 

\BfPara{Significance} Detecting malicious shell commands to harden the security of a device is of paramount importance. 
While the prior works have studied the malicious use of Windows {\em PowerShell}, the malicious use of the Linux shell for attack automation in IoT devices is not fully-investigated. 
This work aims to study shell commands that appear in the static analysis of IoT malware binaries, and understand their intrinsic features towards their detection. It is important to note that there has been some work on understanding shell commands and their use by malicious software in the literature. However, the majority of the prior work has focused on other shell interpreters (\eg power and web), and the emergence of Linux-based IoT malware that heavily uses shell commands makes the detection of shell commands associated with malicious IoT software of paramount importance.

\BfPara{Our Approach} To address this threat, in this work we design, implement, and evaluate \ours, a system for detecting malicious shell commands used in IoT malware.  To evaluate \ours, we collect a dataset of residual shell commands from IoT malware samples. 
Our preliminary analysis shows that shell commands can be found embedded in the disassembled code of malware binaries. 
Therefore, we employ static analysis to search through the disassembled code of malware to extract the shell used in the malware samples. 
For the shell commands were initiated by a benign process, we collect a dataset from benign applications and users. In particular, we use the traffic generated from applications in a real-world environment.  For analyzing and detecting malicious commands, \ours{} employs a Natural Language Processing (NLP) approach for feature generation, followed by deep learning-based modeling for detecting malicious commands.

\BfPara{Contributions}
This work aims to utilize static analysis to detect the malicious use of shell commands in IoT binaries, and to use them as a modality for IoT malware detection. As such, we make two broad contributions. {\bf C-1:} Using shell commands extracted from 2,891 recent IoT malware samples along with a benign dataset, we design a detection system that can detect malicious shell commands with an accuracy of more than 99\%. 
    Compared to the state-of-the-art approaches, 
    our system is more efficient and accurate. 
    Using term- and character-level features, the feature space on the shell commands is easy to explain and interpret. Features contributing to malicious behaviors can be easily identified so that shell commands could be restricted to legitimate use. {\bf C-2:} We extend our command-level detection approach and design a detection model for malicious files (malware samples), which often include multiple commands. Extending the results of detecting individual commands, we group the commands by file and detect the malicious files with an accuracy of more than 99\%. Our detection approach can be applied to files compiled for any processor architecture (\eg ARM, MIPS, Power PC, \etc) as long as the shell commands are extracted, which can be done efficiently. 

\BfPara{Organization} The rest of this paper is organized as follows.  In \autoref{sec:prelim}, we present the problem statement and a high-level overview of our approach. In \autoref{sec:detect}, we review our approach in details; the feature extraction respecting various specifics of the application domain, learning algorithms, and representations. In \autoref{sec:eval}, we review the evaluation of our approach; heuristics developed for extracting shell commands from malicious binaries and benign use contexts, evaluation metrics and settings, and results. In \autoref{sec:related} we review the related work, and draw concluding remarks in \autoref{conclusion}.

\section{Problem Statement and Approach Overview}\label{sec:prelim}
In this section, we begin by the problem statement and a high-level overview of our approach.

\subsection{Problem Statement}\label{sec:problemstatement}

The problem we tackle in this paper is malware detection using shell commands. Given the modality of the analysis of interest, we are also interested in determining whether a given shell command extracted from a binary or a use context is malicious or benign. We approach this problem systematically by modeling shell commands that appear in the residual artifacts of IoT malware binaries. 

The shell command classification problem is formally defined as follows. First, let $\{x_i, y_i\}_{i=1}^N$ be a training set, where $x_i\in\mathbb{R}^d, y_i\in\{0,1\}$; that is, $x_i$ is a feature representation of a shell command $c_i$, where the representation has $d$ real-valued features, and $y_i$ is the corresponding label of ``zero'' if $x_i$ is a shell command initiated by a benign process, and ``one'' otherwise. The classification problem of shell commands is formulated as finding a set of parameters that make up a function $f$ such that $f(x_i)=\hat y_i$ where $||y_i-\hat y_i||$ for all $i$ is minimized (i.e., minimal prediction error). The transformation of $c_i$ into $x_i$ is called feature extraction, denoted by $\Phi(c_i)=x_i$, and is a central contribution of this work through character- and word-level representations. We use those two approaches for their prevalence in representing text and text-like data, which is the case of shell commands. 

The malware detection problem is defined as an extension of the shell command-level classification problem. For that, we use a combined set of shell commands associated with each malware sample as a representation to conduct malware detection. The same definition above is extended to malware $s_i$, where $s_i$ is a collection of shell commands $c_i^j$, for $j=1,\dots,k$, where $x^j$ is the corresponding feature representation of the malware sample $s_i$. Note that the same function $\Phi$ can be extended for the feature representation (e.g., the features associated with the different commands extracted from the same binary sample can be stacked to represent the binary). Similarly, the function $f$ is defined for the binary-level from the command-level classification.

\subsection{A High-level Overview of Our Approach}
The shell is a single point of entry for malware to launch attacks. As such, detecting malicious commands before they are executed on the host will help secure the host. Even though the malware aims to exploit a vulnerability in the device to access its shell, detecting the malicious commands will help mitigate such exploits. 
Our analysis highlights the use of shell commands for infection, propagation, and attack by malware. The Linux capabilities of embedded IoT devices give adversaries the required power to abuse the shell.

\BfPara{Objectives} The main objective of \ours{} is to effectively detect malicious IoT binaries (files) based on their usage of the shell commands.
Upon detecting individual malicious shell commands (i.e., shell commands associated with malware samples), it will be natural to extend the detection to malicious binaries (files) as a whole. 
Thus, we break down the problem into two parts -- 1) detecting malicious commands and 2) detecting malicious files. 

\BfPara{High-Level Design} Our design operates on various binaries of malicious and benign IoT programs. The key idea of \ours is to employ static program analysis tools to extract meaningful representations that can be used eventually to distinguish benign and malicious binaries. To do so, we start with (potential) IoT malware samples and disassemble them to extract shell commands. We establish various heuristics for extracting those commands, and we outline those heuristics in section~\ref{sec:eval}. We repeat the process for (potentially) benign samples as well to explore the power of our representation for malware detection. To make the processing of these commands computationally tractable, we embed those commands into a representation space by extracting term- and character-level feature representations from them using the bag-of-words technique, which is commonly used in NLP tasks. 
Along with the bag-of-words, we use the $n$-grams to represent the commands as feature vectors. Given that those representations may result in high dimensional data representation, we employ the Principal Component Analysis (PCA) for feature reduction before implementing the classification over commands (see  section~\ref{sec:problemstatement} for problem statement). 

\begin{figure}[t]
\centering
\includegraphics[width=0.49\textwidth]{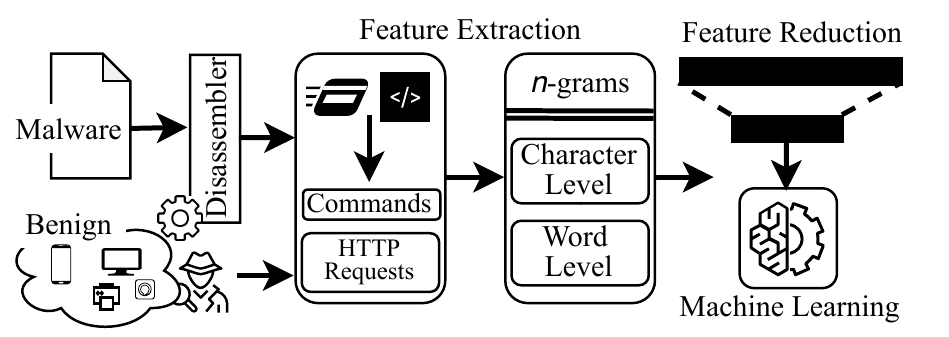}
\caption{The workflow of \ours{}, highlighting the sources of data and its division by class (malicious or benign). The raw data is preprocessed to extract shell commands. The shell commands are represented as (1) characters and (2) words, which are fed to learning networks for detection.
}
\label{fig:system}
\end{figure}

Upon representing the malicious and benign commands as feature vectors, \ours aims to detect malicious commands, as shown in \autoref{fig:system}. To do so, \ours employs machine learning algorithms to classify commands. We use both simple and more advanced (deep) learning approaches. For evaluation, we use cross-validation to address bias and to ensure the generalization capabilities of the model. 
Using the same model architecture, we extend the detection system to detect malicious IoT binaries. To do so, we group the commands by each malware sample and benign application in one single set that is represented as one feature vector to be classified. 

\section{Our Detection System: \ours}\label{sec:detect}
The core of our detection system is a deep learning model built on top of NLP-based features. To better help learn the specifics of shell commands, we tune the default NLP algorithms to enrich the feature representations of the commands. We represent the commands as feature vectors using the bag-of-words approach. Then, we reduce the feature space using PCA. ML-based algorithms are then used for malicious command and sample detection. In the following, we review the technical details of the feature extraction, and classification methods of our detection system.

\subsection{Feature Extraction and Reduction}
The feature extraction process aims to present the attributes of samples, by cleansing and linking the data and transforming it into a format that is easier to process by the employed algorithms for detection. In this section, we discuss selecting features that better represent the characteristics of the samples in the dataset. There are many methods of feature extraction depending upon the nature of the data. Considering the textual nature of our samples, we focus on text-based representation methods. Towards this, we leverage the term-level NLP-based approach by considering words in the samples as features. Additionally, since such an approach misses very crucial attributes, we then employed a character-level NLP approach to meet our goals. 

\subsubsection{Term-level NLP-based model}
We leverage NLP for feature generation, by considering independent words as features and occurrence of space and/or characters as tokenizers, while words with a length greater than two are considered in the bag-of-words for feature vector creation. We adopt the bag-of-words approach, along with $n$-grams. 
Let $I_1$ be the words in a command, and $N$ is the total number of words in the command. Therefore, each word in the command can be represented as $I_{1i}$, where $i \in [1,N]$, such that $I_{1} = {I_{11}, I_{12}, I_{13}, ..., I_{1N}}$.

\subsubsection{Character-level NLP-based Model}
The term-level NLP-based approach does not take the operational symbols, such as the logical operators, in a command into consideration, which undermines many discriminating and dominant characteristics of the shell command, thereby not representing the commands accurately. 
The presence of many shell commands utilizing keywords $l\leq 2$ call for building a more accommodating feature generation mechanism. To do so, we changed the boundaries of the definition of a word by considering every space, special characters, alphabets, and numbers as words, along with the $n$-grams and command statistics. This augments our vocabulary with more granular features to capture the attributes precisely. 
Let $I_2$ be a representation of each character, alphabet, number, \etc, constituting a command, and $N$ is the total number of such constituents in the command. Therefore, every such constituent in the command can be represented as $I_{2j}$, where $j \in [1,N]$, such that $I_{2} = {I_{21}, I_{22}, I_{23}, ..., I_{2N}}$.

\subsubsection{Feature Representation}
To represent every element in the dataset from a defined reference point, they are represented with respect to axes in space. In particular, every command/sample in the dataset is represented as a feature vector in the defined feature space. 
We begin by finding the feature space to determine the dimensionality of the vectors. Particularly, the commands are augmented such that every feature of the commands in the dataset has a representation in the feature space. Every command in the dataset is then represented in a space of $n$ axes, where $n$ is the size of feature space. To do so, we devise multiple representations of the commands, such as including the words in the commands and splitting the commands by spaces and every special character. We also form a feature vector by considering each and every letter and special character as corresponding features in the formed vector. We implemented the bag-of-words method to define our feature space.
The rest of this section explains our feature representation mechanism. 

\subsubsection{Bag-of-Words as Command Embedding}
We generate a representation of commands/samples using the bag-of-words technique. Depending upon the splitting pattern of the samples, we create a central vector that stores all words in the samples. Each sample in the dataset is then mapped to an index in the sparse vector representation, \ie feature vector for every elements in the dataset, where the vector has an index for every word in the vocabulary--- The final vector is represented as the occurrence of each word from the vocabulary in a given command (\ie {\em multi-hot encoding}).

Specifically, to generate the vector representation of each command, or sample, we first created a corpus consisting of the character-level and term-level combinations, referred to as tokens, occurring in either malware or benign commands. The vector, for each command or sample, reflects the frequency of each token in the shell command or sample, respectively.

\subsubsection{Encoding Syntax} An important characteristic of the commands is their syntax. This syntax depends on the structure of the command. Therefore, in addition to the standard features gathered from the commands, we also augment the feature space with feature proximity, to capture the structure of the commands. To do so, we also include the features of $n$-grams. Every $n$ contiguous words in a sample's shell commands are considered as a feature. When using $n$-grams as features, every $n$ contiguous words occurring in a sample are added to the bag of words corresponding to them in the feature space.

For each of the two models, as aforementioned, we create a separate bag of words, such that, the bag contains all the words $I_{1i}^k$, where $i \in [1,N]$ and $k \in [1,m]$, such that $N$ is the total number of words in a command and $k$ is the total number of commands in the dataset. along with the $n$-grams. Therefore, the words in all the commands as per the term-level NLP model, can be combined as
$I_{11}^1, I_{11}^2, I_{11}^3, .... I_{12}^1, I_{13}^1, .... I_{1N}^m$
Let $B$ be the bag of word for the dataset, such that $B = B_1, B_2, B_3, ..., B_t,$ where $t \leq m*N$ and $B_p$, such that $p \in [1,t]$, is unique in $B$. Moving forward, each command $I_i$, where $i \in [1,N]$, can be represented as a feature vector ($F$) with respect to the bag of words $B$, such that the $t^{th}$ index be represented as the frequency of occurrence, of the $t^{th}$ word in the bag, in the command.
$F = {f_{B_1}, f_{B_2}, f_{B_3}, ..., f_{B_t}},$ such that $f_{B_p}$, where $p \in [1,t]$, depicts the frequency of the word, appearing at index $p$ in the bag $B$, in the command $I_i$.

\subsubsection{Feature Reduction}
We capture as many features as possible to achieve accurate results. However, beyond a certain point, the model may suffer from the curse of dimensionality, which causes the performance of the model becomes inversely proportional to the number of features. The usage of a wide variety of features to represent samples leads to a high dimensional feature vector which leads to (i) high cost to perform learning and (ii) overfitting, \ie the model may perform very well on the training dataset, but poorly on the test dataset. 

Dimensionality reduction or feature reduction is applied with the aim of addressing the two problems. We implement PCA for feature reduction to improve the performance and the quality of our classifier of \ours, where the PCA features (components) are extracted from the raw features. PCA itself is a statistical technique used to extract features from multiple raw features, where raw features are of $n$-grams and statistical measurements. PCA creates new variables, named Principal Components (PCs). PCs are linear combinations of the original variables, where a possible number of correlated variables are transformed into a low dimension of uncorrelated PCs (thus the quality improvement). PCA normalizes the dataset by transforming them into a normal distribution with the same standard deviation~\cite{chiang2000fault}, resulting in a  standard representation of variables in order to identify a subset that can best characterize the underlying data~\cite{uguz11}.  

We reduce the $d$-dimensional vector representation of commands to $q$ number of principal components onto which the retained variance under projection is maximal.

\subsection{Classification Methods} 
After representing each sample as a feature vector, we classify them into malicious and benign by leveraging the ML-based algorithms.

\BfPara{Logistic Regression (LR)}
LR is a statistical method that employs a logistic function to model a binary dependent variable, referred to as binary classification (``0'' or ``1''). Given ($X,Y$) as an input training set, LR learns to differentiate between positive (``1'') and negative (``0'') segments for each category,  with the assumption that they have a linear relationship. LR, in the higher domain, estimates and optimizes the boundary between the positive and negative classes by minimizing the following function:
\begin{equation}
    \text{\sf Loss}(f(X),Y) = \begin{cases} 
    -\log(f(X)),~Y = 1\\
    -\log(1-f(X)),~Y \neq 1
    \end{cases}, 
\end{equation}
where $f(X)$ is the LR model's current prediction and $Y$ are the labels of the ground truth set.

\BfPara{Random Forest (RF)}
RF is a non-linear classification algorithm that consists of $N$ decision trees each of which is trained on a collection of random features. The RF method reduces the variation in the performance of individual trees and minimizes the impact of noise on the training process. The final prediction of RF classifier with $N$ decision trees is determined by a majority vote over the predictions or by averaging the prediction of all trees, determined as follows:
\begin{equation}
    f_{RF} = \frac{1}{N}\sum^{N}_{n = 1}f_{n}(X_{s}^'),
\end{equation}
where, for a randomly selected feature set, ($X_{.}^' \subset X_{.}$), $f_{n}$ is the $n^{th}$ tree’s prediction and $X_{s}^'$ is the segment’s $s$ vector.

\subsubsection{Deep Neural Networks (DNN)}
DNN is a type of connected and feed-forward neural networks with multiple hidden layers between the input and output layers. The hidden layers consist of a number of parallel neurons, connected with a certain weight to all nodes in the following layers to generate a single output for the next layer.
Given a feature vector $X$ of length $q$ and target $y$, the DNN-based classifier learns a function $f(.) : R^q \xrightarrow{} R^o$, where $q$ is the input's dimension and $o$ is the output's dimension.
With multiple hidden layers, the dimension of the output of every hidden layer decreases with transformation. Each neuron in the hidden layer transforms the values of the preceding layer using linearly weighted summation, $w_1 + w_2 + w_3 + ... w_q$, which passes through a ReLU activation function ($y(x) = max(x,0)$). The output of the hidden layers is then fed to the output layer, and passed to a softmax activation function $h$, defined as $h(x) =  \frac{\mathrm{1} }{\mathrm{1} + e^{-x}}$, outputting the prediction of the classifier.

\subsection{Term- and Character-level NLP-based Approaches}

\subsubsection{Term-level NLP-based Model}
The term-level learning model uses words as features, with spaces and other special characters as tokenizers. Additionally, it does not consider words less than three characters long. To better represent the locality of the words, the model utilizes $n$-grams. Particularly, it uses 1- to 5-grams. With 10-fold cross-validation.

\subsubsection{Character-level NLP-based model}
We note that the term-level considers the words and neglects the characters, spaces, and words that have a length of less than three. This, in turn, presents a major shortcoming, since a large number of command keywords have a length of fewer than three characters, including {\em cd} and {\em ls}, or consist of special characters, such as {\tt ||} and {\tt \&\&}. 
To address the shortcoming, we create the feature generation step considering these important domain-specific characteristics that would otherwise be ignored. To do so, we change the way in which a word is defined by carefully declaring the tokenizers such that no character is ignored. Subsequently, the changed bag of words considers the character-level, and contains every letter, number, and character represented as an individual feature. 

\section{Evaluation and Discussion}\label{sec:eval}
We divide our evaluation into two parts. First, we build a detection system to detect malicious commands by considering every individual command in the dataset. Second, this detection system is then extended for detecting malicious files, where the above commands corresponding to an application are combined together when representing a single file as a feature vector of multiple commands.

We provide further details of the datasets and their characteristics, and the utilized evaluation metric. We then describe the term-level and character-level NLP-based models. Finally, we describe how these two models are leveraged for detecting individual commands and malicious files.

In addition to the placement of the letters, characters, and spaces, we also consider combinations of these elements in the form of $n$-grams (up to 5-grams) into a vector space. Finally, for feature reduction, we use PCA such that the feature representations preserve 99.9\% of the variance in the training dataset. 

To set out, we begin by describing the process of 
assembling the dataset used in this evaluation. 
We obtain our shell commands by statically disassembling the malware binaries and extracting shell command strings (following some regular expression rules).

\subsection{Malicious Dataset and Commands Extraction}\label{sec:maldata}
We obtain a dataset of 2,891 randomly selected IoT malware samples from the IoTPOT project~\cite{PaSYM2016}, a honeypot emulating IoT devices. 
IoTPOT emulates services, such as telnet and other vulnerable services including those of specific devices with distributed proxy sensors in several countries~\cite{IotpotWebsite21}. Additionally, IoTPOT covers eight different architectures.
Table \ref{tab:arch} depicts the malware distribution according to their architectures and their percentage.
\autoref{fig:system} shows our approach, end-to-end, split into three modules: initial discovery, command extraction, and detection. Our data collection is represented in the first two modules. In the following, we outline the steps we have taken in order to obtain the shell commands from the malware samples (binaries). 

In the initial discovery module, we disassemble the malware binaries. To create a set of rules that automatically apply to samples for retrieving the relevant commands, we manually examine all shell commands extracted from the strings of {\bf 18 malware samples} and establish patterns of those commands.  We then use them to automate the extraction of shell commands for the rest of the malware samples. 

The second component in our workflow is a command extraction module, which takes the command patterns obtained in the initial discovery phase and applies those patterns to the strings of each sample. As a result, 
we extract the shell commands from the malicious binary samples, by concentrating on the {\em strings} only, and label them as malicious. 

\begin{table}[t]
\centering
\caption{Malware dataset by architecture. Percentage is out of the total samples.}
\label{tab:arch}
\begin{tabular}{|l|r|r|}
\Xhline{3\arrayrulewidth}
{\bf Architecture} & {\bf Samples} & {\bf Percentage}\\ 
\Xhline{2\arrayrulewidth}
ARM	&	668	&	23.11\%	 \\ 
MIPS	&	600	&	20.75\%	\\ 
Intel 80368	&	449	&	15.53\%	\\ 
Power PC	&	270	&	9.34\%	\\
X86-64	&	242	&	8.37\%	\\
Renesas SH	&	233	&	8.06\%	\\ 
Motorola m68k	&	217	&	7.51\%	\\ 
SPARC	&	212	&	7.33\%	\\  \Xhline{2\arrayrulewidth}
Total	&	2,891	&	100\%	\\ 
\Xhline{3\arrayrulewidth}
\end{tabular}
\end{table}

\BfPara{Commands Extraction} Using {\em Radare2}, an open-source static analysis tool with an API for automation, we first disassemble each malware binary in our 2,891 samples and extract the {\em strings} from the disassembled code. 
We then use the strings appearing in each sample to obtain the shell commands in them, creating our malicious commands. 
For coverage, we gather all {\em strings} from the disassembled code. 
For a faster extraction of the shell commands, we calculate the offset, or memory address where the string is referenced in the disassembled code, then conduct the disassembly from that offset. 
We pull the instruction set at the offset and extract the desired command.  
Before automating the command extraction, we manually analyze the 18 samples to observe patterns that could uniquely identify the shell commands.

From these 18 malware samples, we identify 1,273 patterns and use them to extract the shell commands from other samples. Our definition of shell commands covers the tasks that can be instructed using a terminal, such as the Linux/Unix-like system commands, HTTP messages, and automated tasks.
For example, strings beginning with shell command keywords, such as {\em cd }, between {\em if} and {\em fi}, {\em kill}, {\em wait}, {\em disown}, {\em suspend}, {\em fc}, {\em history}, {\em break}{}, {\em GET}, {\em POST}, among other similar command structures, are extracted. Malware samples use the shell commands to achieve their objectives, such as traversing  directories (cd), killing a running process of interest (kill), communicating with a C2 (GET, TFTP), and exfiltrating data (POST).
For coverage of those patterns, we use online resources to build a dataset of the keywords of shell commands to augment our automation process.

Based on the identified patterns, we use regular expressions to search for the specific patterns in the {\em strings} obtained from the malware to automate the process for all malware samples. Although the commands contained in the strings may not be syntactically correct, \eg spaces are masked with special characters or spaces, they, however, hint to the location of shell command references. 
We then navigate to the address where a particular string is quoted and disassemble at that offset.

\subsection{Benign Dataset and Commands Extraction}\label{sec:bendata}
To evaluate \ours{}, acquiring a benign shell commands dataset is a necessary step, although a challenging task for multiple reasons. For example, while Linux-based applications are ubiquitous, extracting the corresponding shell commands and using them as a baseline for our benign dataset might be only partially representative, since these binaries may not be necessarily intended for embedded devices. 

Another approach to collect benign shell commands is by observing shell access and their usage by benign users, which requires monitoring network traffic to ``sniff'' the shell commands by benign users. However, we notice that a majority of the traffic nowadays is carried over HTTPS, the encryption limits our visibility into those benign shell commands. 

To cope with these shortcomings, we rely on volunteers for providing their usage of shell commands as a representative of benign usage. In order to do so, we conduct collection efforts at both the host and network levels. 
At the host-side, we gather the bash history data from nine volunteer users. To protect the users' privacy, we anonymize their identities by manually observing the commands and removing every clearly identifying information, such as usernames, domain names and IP addresses, in a consistent manner. 
In total, we collect a dataset of about 143 MB from these volunteers, consisting of 5,772 commands. The collected commands correspond to services, such as {\em ssh}, {\em git}, {\em apt}, {\em Makefile}, and {\em curl}, among others, and generic Linux commands, such as {\em cd}, {\em rm}, {\em chmod}, {\em cp}, and {\em find}, among others.

\begin{figure}[]
\centering
\includegraphics[width=0.40\textwidth]{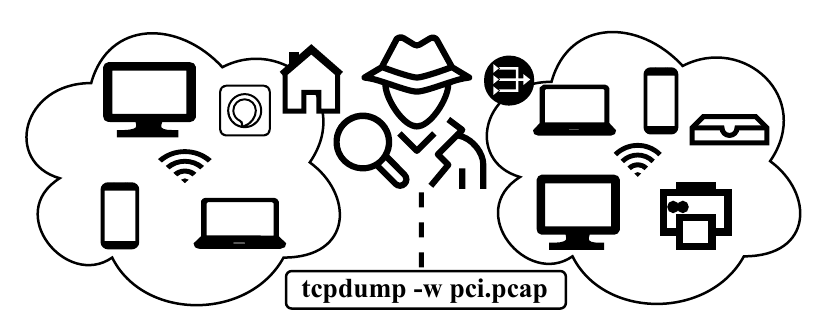}
\caption{Monitoring stations for creation of benign dataset creation. Two network implementations are used: NAT, and a home network.}
\label{fig:NetworkExp}
\end{figure}

For the network-side profiling, we rely on high-level network traffic monitoring from {\bf two networks} to obtain network-level artifacts (\eg GET, POST, etc.) that are not part of an encrypted payload. 
In particular, we look for commands coming from various Linux-based tools, frameworks, and software inject. Since an entry point for many malware families is the abuse of many application-layer protocols, such as HTTP, FTP and TFTP, with the intent to distribute malicious payloads and scripts, we attempt to monitor those protocols in benign use setup for benign data collection.
As such, we built our benign command collection framework with two separate networks, as highlighted in \autoref{fig:NetworkExp}. 

The first network is hidden behind a NAT and consists of five stations, while the second network is a home network with 11 open ports: 21, 22, 80, 443, 12174, 1900, 3282, 3306, 3971, 5900, and 9040. The main purpose of this setup is to capture the incoming and outgoing packets from the home network. Our home network in this experimental setup consists of two 64-bit Linux devices, one Amazon Alexa, one iPhone device, one Mac device with a voice assistant, Siri, which is continuously used, and a router.
\autoref{fig:NetworkExp} is a high-level illustration of our benign data collection system. In the first network (right), we have five devices that are used in a lab setting under ``normal execution'', \ie for everyday use. The network is monitored over a period of 24 hours, where all network traffic is captured.  

The second network is a home network designed by selecting a variety of devices, also operating under ``normal execution'' with the exception that the configured voice assistants in the second network are actively queried during the monitoring time. 
To establish a baseline, the network is monitored without the devices and as the devices are added gradually to the network. 
For the voice assistants, we iterate over a set of questions requiring access to the Internet and actively monitor the traffic at the router for seven hours. Using these settings, we gather a dataset of approximately 34 GB from the first network and approximately 1 GB from the second network.

The traffic gathered from the five volunteers (with consent) in the first network (Network 1) result in a total of 28,578,754 individual payloads, and only 1,625,143 of them are not encrypted, which we utilize for our benign dataset. From the second network (Network 2), five sources generate 4,735 unencrypted payloads in total, which we use as part of our dataset. In total, our benign dataset consists of three parts, {\em bash} (5,772 commands), {\em network 1} (1,625,143 commands), and {\em network 2} (4,755 commands).

\autoref{tab:SampleCommands} shows samples of the payloads from the four data sources. We analyze the samples to find the architecture for which they are compiled using the Linux {\em File} command. In the data representation, we first consider a corpus compiled from both the malware and benign commands to extract the vector representation of each sample. However, determining what is benign is an open challenge, particularly in the domain of malware detection using machine learning, where benign is assumed to not have a fixed pattern, while the malicious samples share behavioral patterns with one another. Toward this, we also investigate using malware-only commands corpus in the process of extracting the vector representation of the samples. This will reduce the bias towards the benign dataset, and demonstrate the effectiveness of the proposed approach, as the feature representation is  generated solely using the malware samples, and is only reflected on the benign samples.

\begin{table*}[t]
\centering
\caption{Data sources in our dataset. ``Sources'' is the number of files used to extract commands, while ``Commands'' is the total number of  commands obtained from the source files.}
\label{tab:SampleCommands}
\begin{tabular}{|l|r|r|l|}
\Xhline{3\arrayrulewidth}
{\bf Data} & {\bf Sources} & {\bf Commands} & {\bf Example} \\ 
\Xhline{2\arrayrulewidth}
{\bf PCAP Net. 1}	&	5	&	1,625,143 &	 GET /update-delta/hfnkpimlhhgieaddgfemjhofmfblmnib/5092/5091/193cb84a\\  & & & 0e51a5f0ca68712ad3c7fddd65bb2d6a60619d89575bb263fc5dec26.crxd HTTP/1.\\ & & & 1\textbackslash{r}\textbackslash{n}Host: storage. googleapis.com\textbackslash{r}\textbackslash{n}Connection: keep-alive\textbackslash{r}\textbackslash{n}User-Agent: \\ & & &Mozilla/5.0 (X11; Linux x86_64) AppleWebKit/537.36 (KHTML, like Gecko) C\\ & & &hrome/72.0.3626.121 Safari/537.36 \textbackslash{r}\textbackslash{n}Accept-Encoding: gzip, deflate\textbackslash{r}\textbackslash{n} \\ 
\hline
{\bf PCAP Net. 2}	&	5 	&	4,735 &   GET /favicon.ico HTTP/1.1\textbackslash{r}\textbackslash{n}Connection: close\textbackslash{r}\textbackslash{n}User-Agent: Mozilla/5.0\\ & & & (compatible; Nmap Scripting Engine; https://nmap.org/book/nse.html)\textbackslash{r}\textbackslash{n}\\ & & &Host: 192.168.2.1\textbackslash{r}\textbackslash{n}
	\\ \hline
{\bf Bash cmd.} & 9 & 5,772 & sudo wget https://download.oracle.com/otn-pub/java/jdk/8u201-b09/4297048\\ & & &7e3af4f5a a5bca3f542482c60/jdk-8u201-linux-x64.tar.gz \\ \hline
{\bf Malware}	&	2,891	&	178,261 &  GET /cdn-cgi/l/chk_captcha?id=\%s \& g-recaptcha-response=\%s HTTP/1.1 \\ & & &User-Agent: \%s Host: \%s Accept: */\* Referer: http://\%s/ Upgrade-Insecure-\\ & & &Requests: 1 Connection: keep-alive Pragma: no-cache Cache-Control: no-cache	\\
\Xhline{3\arrayrulewidth}
\end{tabular}
\end{table*}

\begin{table*}[t]
    \centering
     \caption{Size characteristics of the different datasets. Len. stands for length.}\label{tab:commandLength}
    \begin{tabular}{|l|c|c|c|c|c|c|}
    \Xhline{3\arrayrulewidth}
    \multirow{2}{*}{{\bf Dataset}} &  \multirow{2}{*}{{\bf Commands}} & \multicolumn{5}{c|}{{\bf Command Length Statistics}} \\ 
    \cline{3-7}
          & & {\bf Maximum} & {\bf Minimum} & {\bf Average} & {\bf Median} & {\bf Standard deviation}\\
         \Xhline{2\arrayrulewidth}
         {\bf Network 1} & 1,625,143 & 1,564 & 52 & 184.68 & 185 & 4.88 \\
         {\bf Network 2} & 4,755 & 1,536 & 8 & 209.01 & 167 & 146.26 \\
         {\bf Bash} & 5,772 & 356 & 2 & 23.00 & 14 & 27.71 \\
         {\bf Malware} & 178,261 & 984 & 5 & 293.91 & 384 & 168.03 \\
         \Xhline{3\arrayrulewidth}
    \end{tabular}
\end{table*}

\subsection{Evaluation Settings and Metrics} To evaluate \ours{}, we use the dataset highlighted in \ref{sec:maldata} and \ref{sec:bendata}. In the following, we review settings, parameters tuning, validation technique, and evaluation metrics. 

\subsubsection{Dataset}\autoref{tab:commandLength} shows the number of commands as well as the commands' length statistics (maximum, minimum, average, median, and standard deviation). We notice that commands in Network 1 have similar lengths, as indicated with the low deviation. We notice that Network 2 (corresponding to the IoT devices setting) and Malware datasets have the closest lengths overall, per the average and standard deviation characteristics of their distributions.

\subsubsection{Parameters Tuning} For a better features representation, we utilize $n$-grams. Particularly, we use 1- to 5-grams. For the DNN-based classifier, we also try multiple combinations of parameters to tune the classifier for better performance. We achieve the best performance using five hidden layers. 

\subsubsection{K-Fold Cross-Validation}
To generalize the evaluation, cross-validation is used. For K-fold cross-validation, the data are sampled into K subsets, where the model is trained on one of the K subsets and tested on the other K-1 subsets. The process is then repeated, allowing each subset to be the testing data while the remaining nine are used for training the model. The performance results are then taken as the average of all runs. In this work, We use 10 for K.

\subsubsection{Evaluation Metrics} For a class $C_i$, (where $i \in \{0,1\}$), False Positive (FP), False Negative (FN), True Positive (TP), and True Negative (TN) are defined as: 
\begin{itemize}
    \item TP of $C_i$ is all $C_i$ instances classified correctly 
    \item TN of $C_i$ is all non-$C_i$ not classified as $C_i$ 
    \item FP of $C_i$ is all non-$C_i$ instances classified as $C_i$
    \item FN of $C_i$ is all $C_i$ instances not classified as $C_i$. 
\end{itemize}

We used the Accuracy (AC), F-score (F-1), False-Negative Rate (FNR), and False-Positive Rate as evaluation metrics, which are defined as follows:
\begin{itemize}
    \item $\text{AC}=\text{(TP+TN)}/\text{(TP+TN+FP+FN)}$,
    \item $\text{F-1} = \text{2TP} /  \text{(2TP + FP + FN)}$,
    \item $\text{FNR}=\text{FN}/\text{(TP+FN)}$, 
    \item and $\text{FPR} = \text{FP} /  \text{(FP + TN)}$.
\end{itemize}
We report the metrics as mean AC, mean FNR, and mean FPR for the ten folds.

\begin{table*}[t]
\caption{Evaluation results (\%) of malicious commands detection for both term- and character-level representations. Left: the vector representation is generated from commands extracted from both benign and malware. Right: only the malware commands are used to generate the vector representation.}\label{tab:results}
\centering
\scalebox{0.92}{
\begin{tabular}{|l|c|c|c|c|c|c|c|c|c||c|c|c|c|c|c|c|c|}
\Xhline{3\arrayrulewidth}

\multirow{2}{*}{Target} & \multirow{2}{*}{ML} & \multicolumn{4}{c|}{Term-level} & \multicolumn{4}{c||}{Character-level} & \multicolumn{4}{c|}{Term-level} & \multicolumn{4}{c|}{Character-level} \\\cline{3-18}
&  & Acc. & F-1 & FNR & FPR & Acc. & F-1 & FNR & FPR & Acc. & F-1 & FNR & FPR & Acc. & F-1 & FNR & FPR  \\\Xhline{2\arrayrulewidth}
\multirow{3}{*}{Command} & LR & \textbf{99.86} & \textbf{99.86} & \textbf{0.03} & 0.20 & \textbf{99.87} & \textbf{99.87} & \textbf{0.12} & 0.15 & 89.44 & 89.43 & 14.03 & \textbf{8.43} & 98.03 & 98.03 & \textbf{1.79} & 2.08  \\\cline{2-18}
& RF & 99.84 & 99.84 & 0.09 & 0.20 & 99.78 & 99.78 & 0.27 & 0.19 & 84.95 & 84.96 & 19.52 & 12.29 & 98.19 & 98.19 & 2.32 & \textbf{1.49} \\\cline{2-18}
& DNN & 99.85 & 99.85 & 0.08 & \textbf{0.19} & \textbf{99.87} & \textbf{99.87} & \textbf{0.12} & \textbf{0.14} & \textbf{89.52} & \textbf{89.52} & \textbf{13.27} & 8.77 & \textbf{98.24} & \textbf{98.24} & \textbf{1.79} & 1.74 \\\Xhline{2\arrayrulewidth}

\multirow{3}{*}{File} & LR & 97.76 & 97.77 & 0.63 & 3.04 & 99.70 & 99.70 & \textbf{0.07} & 0.42 & 66.37 & 65.83 & 55.82 & 22.62 & \textbf{99.84} & \textbf{99.84} & \textbf{0.07} & 0.21  \\\cline{2-18}
& RF & \textbf{99.08} & \textbf{99.08} & 1.25 & \textbf{0.76} & \textbf{99.91} & \textbf{99.91} & 0.14 & \textbf{0.07} & \textbf{67.48} & \textbf{67.43} & \textbf{49.06} & 24.32 & 99.79 & 99.79 & 0.28 & \textbf{0.17} \\\cline{2-18}
& DNN & 97.16 & 97.18 & \textbf{0.35 }& 4.08 & 99.28 & 99.29 & 0.14 & 1.00 & 66.67 & 65.58 & 55.95 & \textbf{22.10} & 99.26 & 99.26 & 0.14 & 1.04 \\
\Xhline{3\arrayrulewidth}
\end{tabular}}
\end{table*}

\subsection{Detecting Malicious Commands}
We use \ours{} to detect individual malware commands. We first present the results of the term-level model, followed by the character-level model.
On average, the term-level model provides an accuracy of more than 99\% along with an FNR of less than 0.1\% and FPR of less than 0.20\% as shown in \autoref{tab:results} (left), with all approaches performing similarly.
We then test the performance of the character-level NLP-based model for detecting individual malicious commands over the same dataset. 
As shown in \autoref{tab:results} (left), the approach achieved similar results on the term-level, with up to 99.87\% accuracy using LR and DNN.

\BfPara{Malware-based Corpus} Next, we investigate the performance of both term- and character-level representations extracted using the malware samples only, and without considering the benign samples. 
Benign software are diverse in behavior, intuition, appearance, and goal, while malicious software share commonalities within their design~\cite{BarrenoNJT10}. Therefore, identifying benign samples should not be dependent on the existence of ``benign'' patterns, but the non-existence of ``malicious'' patterns. This, in turn, will reduce the bias toward the training benign dataset, which is essential considering that obtaining a representative benign dataset is an open challenge. 
Therefore, the benign samples were excluded from the process of constructing the feature space for this evaluation. \autoref{tab:results} (right) shows the performance of both representations. In contrast to the previous results, the term-level representation performance was significantly affected ($\sim$89.52\% accuracy), while the character-level representation maintains a performance of up to 98.24\% (only 1.6\% performance degradation) using the DNN-based model.

\subsection{Malware Detection}
To generalize from the shell command detection to binaries (malware) detection, we classify files as malicious or benign using vectors of feature per file that combine the feature values of the shell commands associated with each file.

\subsubsection{Dataset} For this task, we generate benign samples, drawn from benign commands randomly selected to follow similar command-frequency distribution as the malicious samples. 
We first generate the command-frequency distribution, \ie defined the distribution of number of commands per sample, of the real-world malicious samples in our dataset. Then by using the sampling techniques, we generate a statistically similar (size-wise) dataset of benign samples that fall in the same size as the malicious samples.

\subsubsection{Model Training and Detection Performance} Subsequently, we train and test the model over the file specific dataset. In doing so, the commands corresponding to a file are represented as a feature vector of that file. Similar to the individual commands detection, as shown in \autoref{tab:results} (left), we evaluated both the term-level and character-level representations, with the character-level model yielding a higher detection rate of 99.91\% with 0.14\% and 0.07\% of FNR and FPR, respectively. Compared to the term-level model, the character-level model performs better and improves the accuracy by $\approx$2\% and also reduces the FPR and FNR. This reflects the improved feature representation technique and emphasizes the importance of special characters.

\BfPara{Malware-based Corpus} We also investigate the performance of malware-only term- and character-level representations for malicious files and commands detection. \autoref{tab:results} (right) show the performance of both representations. Similar to the malware command detection, the term-level approach performance is highly affected, reduced from 99\% to 67\% accuracy. However, the character-level approach maintained similar results (\ie 1\% performance reduction in malware command detection). This indicate the stability of the character-level representations in malicious behavior modeling, a characteristic that may not hold true for the term-level representation.

\section{Discussion}
Prior works have used different approaches for IoT malware detection, including control flow graphs and image-based representation of binaries.
Studies have also shown that the Linux-based malware are structurally different from the traditional Android malware~\cite{AlasmaryA0CNM18}. Further, although studies have shown the abuse of the windows Powershell~\cite{HendlerKR18}, they differ functionally from the Linux shell. Considering the use of shell by the adversaries towards their intent~\cite{AnwarAPWCM20}, we argue that shell commands can be used as a modality for effective detection as well.
Additionally, a shell command-based modality can be used to detect fileless malware. We further discuss the implications of this work in the following.

\subsection{The Use of Shell as a Weapon}
The shell abstracts details of the communication between the application and the Operating System (OS), and is used by applications for interacting with the file system, OS, etc.
However, adversaries use shell commands for their malicious intents, e.g., interacting with the command and control server, directory traversal, and data exfiltration. This can be facilitated by the use of default credentials by the owners and vulnerabilities in the services such as SSH, and device firmware. The vulnerabilities in the firmware could be due to the usage of outdated firmware or due to delayed upgrading of firmware or services. For example, in 2014, Shellshock bash attacks caused a vulnerability in Apache systems through HTTP requests and using the {\em wget} command to download a file from a remote host and save it to the {\em tmp} directory to cause infection~\cite{Koch15}. 

A recent vulnerability (CVE-2019-1656), which results from the improper input validation in Linux operating system and can be exploited by the adversaries by sending crafted commands to gain access to targeted devices, has been reported~\cite{nvd_cvss}. 
By abusing the shell, adversaries can utilize the shell to brute-force the credentials of users to gain access to the device by launching a dictionary attack. Additionally, they can use the shell to connect to C2 servers to download instructions; \eg infecting the device, propagating itself, or launching a series of directed flooding attacks. Moreover, malware can use bash to {\em find} command to look for uninfected files in the host device and use the {\em tmp} directory to download and run malware.

\subsection{Detecting Individual Shell Commands} Although researchers have looked into the malicious usage of Windows {\em PowerShell}, and except for analyzing the vulnerabilities in Linux shell (\eg shellshock), the malicious usage of shell commands has not been analyzed in the past. Prior works have analyzed and detected the use of shell commands to propagate attacks, \eg sending malicious bots~\cite{Geer05}, and installing ELF executables on Android systems~\cite{SchmidtSCYKCA08}. Given the larger ecosystem of connected embedded devices with Linux capabilities, and sensing the urgency, we analyze the usage of shell commands used by malware. We propose a system to detect malicious commands with 99.8\% accuracy.

\subsection{Malware Detection}
Many efforts have been dedicated to address the security threats to IoT from the hardware, the software and the application perspectives.
Some also argue that there is a need for a cross-layer approach for comprehensive protection of the IoT systems~\cite{wang2019xlf}.
Meanwhile, IoT malware has been on the rise. 
Given the difficulty of obtaining samples, very few works have been done on detecting IoT malware, and even less using residual strings in the binaries either.
\autoref{sec:related} discusses the methods that work on detecting IoT malware. 
In this work, we use the commands in the malware samples for detecting them. 
Our detection model achieves an accuracy of 99.8\% with FNR and FPR of 0.2\% and 0.1\%, respectively.
As malware abuse the shell of the host device, detecting them at the shell will help safeguard the device from becoming infected. Additionally, malware access a device by breaking into the host device by launching a dictionary attack, typically a single shell command execution. Alternatively, a host device can also be infected by a zero-day vulnerability or an outdated device with an existing exploitable vulnerability, among others, which are also executed by individual shell commands. For a successful event, where the adversary breaks into a host, it will then abuse the shell to infect the host, followed by propagating the malware, and creating a network of botnets to launch attacks. As such, having a detector of such high accuracy, at both the individual command level and malware sample level, with low FPR and FNR, will help stop the host device from being used as an intermediary target for launching attacks, despite the presence of vulnerabilities or the host. This makes this work very timely and necessary.

\subsection{Applications} ShellCore detects malicious software by leveraging the use of shell in binaries. Given the increase in IoT malware attacks, their use of shell commands can be leveraged for their detecting and the associated malware intent unveiled by the commands. Additionally, ShellCore can be leveraged to detect fileless attacks~\cite{Devine17}, a new and emerging type of attacks where the adversary uses a target device's terminal to execute successive commands that implement the malicious intent. As a file is unavailable for analysis, the execution of commands can be used as a modality towards their analysis and detection.

\subsection{Limitations} In this study, we analyzes the IoT malware statically to extract shell commands from the malware disassembly. Thus, our approach is limited to malware that do not employ obfuscation. Prior studies have shown that obfuscation is still uncommon among IoT malware~\cite{CozziVDSBB20}, making our model applicable under existing circumstances. Additionally, prior studies have also shown the use of standard packers by IoT malware, e.g., UPX~\cite{upx,CozziVDSBB20}. The standard packers's {\em unpack} module can thus be leveraged to extract the malware binary, and our model can then be used to detect the malicious software.

A major challenge in our study is generating a reliable dataset of both malware and benign samples. While we reconstruct the command usage by malware through extracting commands from the malware codebase, we extract the benign usage from the shell by the Linux-based devices. Additionally, the benign dataset may not be considered as a representative ground-truth benign dataset with an absolute confidence. Therefore, and to account for that shortcoming, we evaluated our models using representations extracted exclusively from the malware samples.

\section{Related Work}\label{sec:related}
A summary of the related work is in~\autoref{tab:related}. 
Broadly, there have been some work on {\em PowerShell} and {\em Web Shell} commands detection, as well as IoT malware detection, which are related to this work. No prior work exists on IoT shell commands. 

\begin{table*}[h]
\centering
\caption{Comparison with Related work. AUC: Area Under the Curve, TPR: True Positive Rate, TNR: True Negative Rate, AC: Accuracy, FNR: False Negative Rate, FPR: False Positive Rate, NLP: Natural Language Processing, CNN: Convolutional Neural Networks, MS: Malware Signature,MF:  Malware Functions, LW: Longest Word in files header, DL: Deep Learning, MLP: Multi-Layer Perceptron, SVM: Support Vector Machine, GBT: Gradient Boosted Tree, LSTM: Long Short Term Memory, SDA: Static and Dynamic Analysis, Pr.: Precision, Re.: Recall, and F1: F1-score. Note that our system is capable of classifying both shell commands and hosting malware. *The last row demonstrates the results of our system in classifying malware samples using their shell commands.}
\label{tab:related}
\scalebox{0.99}{
\begin{tabular}{|l|l|l|l|l|l|}
\Xhline{3\arrayrulewidth}
{\bf Study} & {\bf Shell Type}  & {\bf Dataset} & {\bf Capability} & {\bf Performance (Best Result)}   & {\bf Method}  \\ 
\Xhline{2\arrayrulewidth}
Starov~\etal\cite{StarovDAHN16}  & Web shell           & 481     & Analysis   &          
 --- &         \begin{tabular}[c]{@{}l@{}}SDA \end{tabular}                                                                                                                                        \\ \hline

Uitto~\etal\cite{UittoRML15}   & Linux shell         &    13,257     & Analysis   &    
 --- &     Diversification                                                                                                                                           \\ 
\Xhline{2\arrayrulewidth}
Tian~\etal~\cite{TianWZZ17} & Web shell           & 7,681   & Detection  &
 \begin{tabular}[c]{@{}l@{}}Pr. (98.6\%), Re. (98.6\%), F1 (98.6\%)\end{tabular}
 & CNN                                                                                                                                            \\ \hline
Rusak~\etal\cite{RusakAO18} & PowerShell          & 4,079  & Detection  & AC (85\%)                                                                                      & DL                                                                                                                                             \\ \hline

Hendler~\etal\cite{HendlerKR18}   & PowerShell          & 66,388  & Detection  & 
\begin{tabular}[c]{@{}l@{}}AUC (98.5-99\%), TPR (0.24-0.99\%)\end{tabular}   
 & \begin{tabular}[c]{@{}l@{}}NLP, CNN\end{tabular}                                                                                              \\ \hline

Li~\etal~\cite{LiHIMZD19} & PHP web shell          &  950  & Detection  & 
\begin{tabular}[c]{@{}l@{}}AUC (98.7\%), AC (91.7\%), FPR (1.0\%)\end{tabular}
 & RF, SVM, GBT                                                                                                                                             \\  \hline

Stokes~\etal~\cite{StokesAM20} & VBScripts          &  240,504  & Detection  & 
\begin{tabular}[c]{@{}l@{}}TPR (69.3\%), FPR (1.0\%)\end{tabular}
 & LSTM, CNN                                                                                                                                             \\  \hline

Ours (Command-level) & Linux shell          & 190,897   & Detection  & 
\begin{tabular}[c]{@{}l@{}}AC (99.89\%), FNR (0.08\%), FPR (0.13\%)\end{tabular}
 & DNN, SVM                                                                                                                                             \\ 
\hline
Ours (Binary-level) & Linux shell          & 2,891*   & Detection  & 
\begin{tabular}[c]{@{}l@{}}AC (99.83\%), FNR (0.13\%), FPR (0.20\%)\end{tabular}
 & DNN, SVM                                                                                                                                             \\ 
\Xhline{3\arrayrulewidth}
\end{tabular}}
\end{table*}

\subsubsection{Shell Commands}
Hendler~\etal\cite{HendlerKR18} detected malicious {\em PowerShell} commands using several machine learning approaches, \eg NLP and Conventional Neural Network (CNN).
Both studies have focused on shell commands that can only run on Microsoft Windows, \ie handling binaries of a single architecture, with very little insight of whether the approach can be applied to IoT software and command artifacts. 
Additionally, Uitto~\etal\cite{UittoRML15} proposed a command diversification technique, by modifying and extending commands, to protect against injection attacks. 
Further, Anwar~\etal~\cite{AnwarAPWCM20} statically analyze the IoT malware and specify about the presence of shell commands in their disassembly. They use them along with other features, such as, strings, Control Flow Graphs (CFG) towards malware detection.

\subsubsection{Web Shell}
Web shell is a script that allows an adversary to run on a targeted web server remotely as an administrator.  Starov~\etal\cite{StarovDAHN16} statically and dynamically analyzed a set of web shells to uncover features of malicious hypertext preprocessor shells.
Tian~\etal~\cite{TianWZZ17} proposed a system to detect malicious web shell commands using CNN and word2vec-based approaches.
In a similar context, Rusak~\etal\cite{RusakAO18} proposed a deep learning approach to classify malicious {\em PowerShell} by families using the abstract syntax trees representation of the {\em PowerShell} commands.
Li~\etal~\cite{LiHIMZD19} propose an ML model to detect malicious web shells written in PHP, achieving an accuracy 91.7\%. 
Moreover, Stokes~\etal~\cite{StokesAM20} employ a recurrent deep learning model to detect malicious VBScripts by using a dataset of first 1000-bytes of 240,504 VBScript files and achieving a TPR of 69.3\% and an FPR of 1.0\%.

\subsubsection{IoT Malware Detection}
IoT malware has been on the rise and has received the attention of researchers which is evident by the growing body of work in this domain.
Pa~\etal\cite{PaSYM2016} proposed IoTPOT, a detection system that supports different malware architectures to analyze and detect Telnet-based attacks on IoT devices.
Dang\etal\cite{DangLLZCXCY19} deployed four IoT honeypots to study the recent fileless attacks launched by Linux-based IoT devices; these attacks do not rely on the malware files and leave no footprint. They found that 99.7\% of fileless attacks use shell commands, making \ours very relevant, since it is capable of detecting these types of attacks. Another work proposed by Perdisci~\etal\cite{PerdisciPAA20} introduced IoTFinder, a multi-label classifier that automatically learns statistical DNS traffic fingerprints for large-scale detection of IoT devices. Alrawi~\etal\cite{AlrawiLAM19} proposed a modeling methodology for home IoT devices to identify unencrypted traffic and other vulnerability.

Recent works have focused on detecting IoT malware traffic, e.g., IoT network packets~\cite{McDermottMP18,KumarL19}, by introducing EDIMA (Early Detection of IoT Malware Network Activity). EDIMA is an IoT malware detection method using supervised ML algorithms atop the analyzed traffic of IoT devices and large-scale network scanning. Bendiab~\etal\cite{BendiabSAK20} proposed a zero-day malware detection and classification method using deep learning atop the analyzed IoT malware traffic and visual representations. The challenge of finding the best suitable algorithm to use in extracting features from executable files was addressed by Darabian~\etal\cite{DarabianDHTAHCP20} using multi-view data extraction. 
Another approach proposed by Liu~\etal\cite{LiuDZZWG19} used a pre-trained random forest that considers the values of misclassification features to build a generic algorithm. Their proposed framework's primary goal is to detect IoT malware on the Android OS with prior knowledge about the devices.

Su~\etal\cite{SuVPSFS18} used a lightweight CNN for IoT’s malware families classification after converting their binaries to grayscale images and achieved 94.0\% of accuracy in classifying DDoS malware in IoT networks and 81.8\% of accuracy in detecting two prominent malware families (i.e., Mirai and Linux Gafgyt). Lei~\etal\cite{LeiQWLY19} introduced a graph-based IoT malware detection technique called ``EveDroid'' as an event-aware Android malware detection tool. Instead of using the API calls to capture malware behavior, EveDroid uses event groups to exploit the apps’ behavioral patterns at a higher level while providing an F1-score of 99\%. In the health-care domain of IoT, Guerar~\etal\cite{GuerarMMP18} discussed  malware vulnerabilities of mobile operating systems and IoT sensors. They introduced the Invisible CAPTCHA to decide if a user is a bot or not by considering the tap and vibration events from the user recorded behaviors while using the mobile devices rather than asking the user to enter the CAPTCHA content manually.

Bertino and Islam~\cite{BertinoI17} proposed a behavior-based approach that combines behavioral artifacts and external threat indicators for malware detection. The approach, however, relies on external online threat intelligence feeds (\eg VirusTotal) and cannot be generalized to other than home network environments (due to computations offloading). 
On the other hand, Hossain~\etal\cite{HossainHZ18} proposed Probe-IoT, a forensic system that investigates IoT-related malicious activities. 
Similarly, Montella~\etal\cite{MontellaRK18} proposed a cloud-based data transfer protocol for IoT devices to secure the sensitive data transferred among different applications, although not addressing the insecurity of the IoT software itself. 
Cozzi~\etal\cite{CozziGFB2018} analyzed a large Linux malware dataset by studying their behavior, and discussed obfuscation techniques that malware authors use. 
Furthermore, Alasmary~\etal\cite{AlasmaryAPCNM19} and Anwar~\etal\cite{AnwarAPWCM20} use the different artifacts of the IoT malware, such as, CFGs, strings, and functions, to build detection systems.
Taking this forward, Abusnaina~\etal examined the robustness of CFG-based IoT malware detection models to adversarial attacks~\cite{AbusnainaKA0AM19} and also proposed effective defenses~\cite{AlasmaryAJAANM20}.
Recent arts have also focused on exploring the IoT network environment. Choi~\etal\cite{ChoiAAWCNM19} explored the presence of endpoints the disassembly of the IoT malware binaries towards characterizing IoT malware spread and affinities. This emphasis on the network has also enabled the monitoring and detection of anomalies and vulnerabilities in wireless communication and network traffics~\cite{JiaXYCLW18,WanXX020,GuFAFHM20}.  

\section{Conclusion}\label{conclusion}
We proposed \ours, a machine learning-based approach to detect shell commands used in IoT malware. 
We analyze malicious shell commands from a dataset of 2,891 IoT malware samples, along with a dataset of benign shell commands assembled corresponding to benign applications.
\ours leverages deep learning-based algorithms to detect malicious commands and files, and NLP-based approaches for feature creation. \ours detects individual malicious commands and malware with an accuracy of more than 99\%,
with low FPR and FNR, when detecting malware. The results reflect that despite a comparatively low detection rate for individual commands, the proposed model is able to detect their source with high accuracy.

\balance
\bibliographystyle{ACM-Reference-Format}
\bibliography{ref,conf}

\end{document}